\documentclass[runningheads, hidelinks]{llncs}
\usepackage{graphicx}
\usepackage{amsmath}  
\usepackage{booktabs}
\usepackage{caption}
\usepackage{makecell}

\usepackage{mathtools}
\usepackage{multirow}
\usepackage{bbold}
\usepackage{hyperref}
\usepackage{xcolor}

\usepackage{pifont}
\usepackage{cite}
\usepackage{paralist}
\usepackage{enumitem,kantlipsum}
\usepackage{caption}
\usepackage{subcaption}
\usepackage{amsmath,amsfonts,amssymb} 

\newcommand{\Plus}[1]{\textcolor{black}{\textbf{\texttt{+}#1}}}
\newcommand{\Minus}[1]{\textcolor{black}{\textbf{\texttt{-}#1}}}

\newcommand{\squishlist}{
 \begin{list}{$\bullet$}
  { \setlength{\itemsep}{0pt}
     \setlength{\parsep}{1pt}
     \setlength{\topsep}{1pt}
     \setlength{\partopsep}{0pt}
     \setlength{\leftmargin}{1.5em}
     \setlength{\labelwidth}{1em}
     \setlength{\labelsep}{0.5em} } }

\newcommand{\squishend}{
  \end{list}  }

\newcommand{\paragraphHd}[1] {\vspace{1.2mm}\noindent\textbf{#1}} 

\begin{document}
\title{A Unified Framework for\\Learned Sparse Retrieval}
%
%
\author{Thong Nguyen\inst{1} \and
Sean MacAvaney\inst{2} \and
Andrew Yates\inst{1}}
\institute{}
\institute{University of Amsterdam \and University of Glasgow \\
\email{t.nguyen2@uva.nl}}
\maketitle              
\begin{abstract}
Learned sparse retrieval (LSR) is a family of first-stage retrieval methods that are trained to generate sparse lexical representations of queries and documents for use with an inverted index. Many LSR methods have been recently introduced, with Splade models achieving state-of-the-art performance on MSMarco. Despite similarities in their model architectures, many LSR methods show substantial differences in effectiveness and efficiency.
Differences in the experimental setups and configurations used make it difficult to compare the methods and derive insights.
In this work, we analyze existing LSR methods and identify key components to establish an LSR framework that unifies all LSR methods under the same perspective. We then reproduce all prominent methods using a common codebase and re-train them in the same environment, which allows us to quantify how components of the framework affect effectiveness and efficiency. We find that (1) including document term weighting is most important for a method's effectiveness, (2) including query weighting has a small positive impact, and (3) document expansion and query expansion have a cancellation effect. As a result, we show how removing query expansion from a state-of-the-art model can reduce latency significantly while maintaining effectiveness on MSMarco and TripClick benchmarks. Our code is publicly available.\footnote{Code: \url{https://github.com/thongnt99/learned-sparse-retrieval}}

\end{abstract}
\section{Introduction}
Neural information retrieval has becoming increasingly common and effective with the introduction of transformers-based pre-trained language models \cite{lin2021pretrained}. Due to latency constraints, a pipeline is often split into two stages: first-stage retrieval and re-ranking. 
The former focuses on efficiently retrieving a set of candidates to re-rank, whereas the latter focuses on re-ranking using highly effective but inefficient methods.
Neural first-stage retrieval approaches can be grouped into two categories: dense retrieval (e.g., \cite{karpukhin2020dense, khattab2020colbert,xiong2020approximate}) and learned sparse retrieval (e.g., \cite{formal2021splade, zhuang2021tilde, zamani2018neural}).
Learned sparse retrieval (LSR) methods transform an input text (i.e., a query or document) into sparse lexical vectors, with each dimension containing a term score analogous to TF.
The sparsity of these vectors allows LSR methods to leverage an inverted index.
Compared with dense retrieval, LSR has several attractive properties.
 Each dimension in the learned sparse vectors is usually tied to a term in vocabulary, which facilitates transparency. We can, for example, examine biases encoded by models by looking at the generated terms. Furthermore, LSR methods can re-use the inverted indexing infrastructure built and optimized for traditional lexical methods over decades.

The idea of using neural methods to learn weights for sparse retrieval predates transformers \cite{zamani2018neural, zheng2015learning}, but approaches' effectiveness with pre-BERT methods is limited.
With the emergence of retrieval powered by transformer-based pre-trained language models \cite{vaswani2017attention,devlin2019bert,lin2021pretrained}, many LSR methods \cite{formal2021splade,formal2022distillation, macavaney2020expansion, lin2021few, dai2020context, mallia2021learning, zhao2020sparta} have been introduced that leverage transformer architectures to substantially improve effectiveness.
Among them, the Splade \cite{formal2021splade} family is a recent prominent approach that shows strong performance on the MSMarco \cite{nguyen2016ms} and BEIR benchmarks \cite{thakur2021beir}. 

 

Despite their architectural similarities, different learned sparse retrieval methods exhibit very different behaviors regarding effectiveness and efficiency. The underlying reasons for these differences are often unclear.

In this work, we conceptually analyze existing LSR methods and identify key components in order to establish a comparative framework that unifies all methods under the same perspective. Under this framework, the key differences between existing LSR methods become apparent.
We first reproduce methods' original results, before re-training and re-evaluating them in a common environment that leverages best practices from recent work, like the use of hard negatives.
We then leverage this setting to study how key components influence a model's performance in terms of efficiency and effectiveness. We investigate the following research questions:

\paragraphHd{RQ1: Are the results from LSR papers reproducible?}\\
 This RQ aims to reproduce the results of all recent, prominent LSR methods in our codebase, consulting the configuration on the original papers and codes. We find that most of the methods can be reproduced with MRR comparable to the original work (or slightly higher).

\paragraphHd{RQ2: How do LSR methods perform with recent advanced training techniques?}\\
Splade models \cite{formal2021splade} show impressive ranking scores on MSMarco.
While these improvements could be due to architectural choices like incorporating query expansion, Splade also benefits from an advanced training process with mined hard negatives and distillation from cross-encoders.
Our experiments show that with the same training as Splade, many older methods become significantly more effective. Most noticeably, the MRR@10 score of the older EPIC\cite{macavaney2020expansion} model was boosted by 36\% to become competitive with Splade.  

\paragraphHd{RQ3: How does the choice of encoder architecture and regularization affect results?}\\
The common training environment we use to answer RQ2 allows us to quantify the effect of various architectural decisions, such as expansion, weighting, and regularization. We find that document weighting had the greatest impact on a system's effectiveness, while query weighting had a moderate impact, though query weighting improves latency by eliminating non-useful terms. Notably, we observed a cancellation effect between improvements from document and query expansion, indicating that query expansion is not necessary for a LSR system to perform well.

\textbf{Our contributions are}: (\textbf{1}) an conceptual framework that unifies all prominent LSR methods under the same view, (\textbf{2}) an analysis of how LSR components affect efficiency and effectiveness, which e.g. leads to a modification that reduces more than 74\% retrieval latency while keeping the same SOTA effectiveness, and (\textbf{3}) implementations of all studied methods in the same codebase, including simple changes in Anserini\cite{yang2018anserini} that make LSR indexing faster.

\section{Learned sparse retrieval}
Learned sparse retrieval (LSR) uses a query encoder $f_Q$ and a document encoder $f_D$ to project queries and documents to sparse vectors of vocabulary size: $w_q=f_Q(q)=w_q^1,w_q^2,\dots,w_q^{|V|}$ and $w_d=f_D(d)=w_d^1,w_d^2,\dots,w_d^{|V|}$. The score between a query a document is the dot product between their corresponding vectors: $sim(q,d) = \sum_{i=1}^{|V|}w_q^iw_d^i$. This formulation is closely connected to traditional sparse retrieval methods like BM25; indeed, BM25 \cite{robertson-1994-okapi,robertson2009probabilistic} can be formulated as:
{\small
\begin{align*}
    \text{BM25}(q,d) &= \sum_{i=1}^{|q|} \text{IDF}(q_i) \times  \frac{tf(q_i, d) \times (k_1 + 1)}{tf(q_i, d) + k_1 \cdot \left(1 - b + b \cdot \frac{|d|}{\text{avgdl}}\right)} \\ 
    &= \sum_{j=1}^{|V|} 
    \underbrace{ \mathbb{1}_{q(v_j)} \text{IDF}(v_j)}_{\text{query encoder}} 
    \times  \underbrace{\mathbb{1}_{d(v_j)} \frac{tf(v_j, d) \times (k_1 + 1)}{tf(v_j, d) + k_1 \cdot \left(1 - b + b \cdot \frac{|d|}{\text{avgdl}}\right)}}_{\text{doc encoder}} \\
    &= \sum_{j=1}^{|V|} f_Q(q)_j \times f_D(d)_j \\
\end{align*}
}%
With BM25 the IDF and TF components can be viewed as query/document term weights. LSR differs by using neural models, typically transformers, to predict term weights.
LSR is compatible with many techniques from sparse retrieval, such as inverted indexing and accompanying query processing algorithms.
However, differences in LSR weights can mean that existing query processing optimizations become much less helpful, motivating new optimizations \cite{mackenzie2021wacky,10.1145/3477495.3531774,10.1145/3576922}.

\begin{table}[ht!]
    \centering
    \small
    \begin{tabular}{lllll}
    \toprule \toprule
    \textbf{Name} & \textbf{Backbone} & \textbf{Head} & \textbf{Expansion}\ \ \ & \textbf{Weighting} \\
    \midrule
        BINARY\ \ \ & Transf. Tokenizer\ \ \ & -  & No & No\\ 
        MLP  & Transf. Encoder & Linear(s) & No & Yes\\ 
       expMLP & Transf. Encoder & Linear(s)  & Yes & Yes\\ 
       MLM  & Transf. Encoder &  MLM Head + Agg.\ \ \ & Yes & Yes \\
       clsMLM & Transf. Encoder & MLM Head & Yes & Yes \\  
    \bottomrule
    \end{tabular}
    \caption{Encoder architectures. (Transf: Transformers)}
    \label{tab:sparse_encoders}
    \vspace{-6mm}
\end{table}

\subsection{Unified learned sparse retrieval framework}
In this section, we introduce a conceptual framework consisting of three components (\textit{sparse encoder}, \textit{sparse regularizer}, \textit{supervision}) that captures the key differences we observe between existing learned sparse retrieval methods.
Later, we describe how  LSR methods in the literature can be fit into this framework. 
\subsubsection{Sparse (Lexical) Encoders.}
A sparse or lexical encoder encodes queries and passages into weight vectors of equal dimension. This is the main component that determines the effectiveness of a learned sparse retrieval method. There are three distinct characteristics that make sparse encoders different from dense encoders. The first and most straightforward difference is that sparse encoders produce sparse vectors (i.e., most term weights are zero). This sparsity is controlled by sparse regularizers, which we will discuss in the next section.

Second, dimensions in sparse weight vectors are usually tied to terms in a vocabulary that contains tens of thousands of terms. Therefore, the size of the vectors is large, equal to the size of the vocabulary; each dimension represents a term (typically a BERT word piece). On the contrary, (single-vector) dense retrieval methods produce condensed vectors (usually fewer than $1000$ dimensions) that encode the semantics of the input text without a clear correspondence between terms and dimensions.
Term-level dense retrieval methods like ColBERT \cite{khattab2020colbert} do preserve this correspondence.

The third distinction is that encoders in sparse retrieval only produce non-negative weights, whereas dense encoders have no such constraint. This constraint comes from the fact that sparse retrieval relies on software stacks (inverted indexing, query processing algorithms) built for traditional lexical search (e.g., BM25), where weights are always non-negative term frequencies.

Whether these differences lead to systematically different behavior between LSR and dense retrieval methods is an open question.
Researchers have observed that LSR models and token-level dense models like ColBERT tend to generalize better than single-vector dense models on the BEIR benchmark\cite{thakur2021beir, formal2022distillation}.
There are also recent works proposing hybrid retrieval systems that combine the strength of both dense and sparse representations \cite{lin2021densifying, lin2022dense, chen2021salient}, which can bring benefits for both in-domain and out-of-domain effectiveness \cite{lin2022dense}.   

There are several variants of sparse encoders, which are typically built on a transformer-backbone\cite{vaswani2017attention} with additional head layer(s) on top.  In Table \ref{tab:sparse_encoders}, we summarize a list of common architectures of sparse encoders proposed in the literature.
We use the following notation when describing these sparse encoder architectures: $v_i$ denotes the $i^{th}$ term in a vocabulary $V$; $t_j$ denotes the $j^{th}$ term in an input sequence $t$ (either a query or document) of length $L$; $h_j$ represents the contextualized embedding of $t_j$ from a transformer encoder; $e_i$ represents the transformer's input embedding of the $v_i$; $w_i(t)$ represents the weight of $v_i$ in the context of $t$.
The architectures include:
\squishlist
    \item \textbf{BINARY}: The BINARY encoder simply tokenizes the input into terms (word pieces) and considers the presence of terms in the input text. The binary encoder performs neither term expansion nor weighting:
    \begin{equation}
    w_i(t) = \max_{j=1..L} \mathbb{1}\big(v_i = t_j)
    \end{equation}
    \item \textbf{MLP}: This encoder uses a \textbf{M}ulti-\textbf{l}ayer \textbf{P}erceptron (usually one layer) on top of each contextualized embedding $h_j$ produced by the transformer-backbone for each input term to generate the term's score. Only terms in the input receive a weight; the other terms are zero. 
    \begin{equation} \label{eq: mlp}
        w_i(t) = \sum_{j=1...L} log\Bigg(\mathbb{1}(v_i = t_j) \bigg(ReLU(h_j W + b)\bigg) +1 \Bigg)
    \end{equation}
    where $W$ and $b$ are the weight and bias of the linear head.
    This MLP architecture focuses on term weighting. 
    \item \textbf{expMLP}: This encoder adds a pre-processing step to expand the input with relevant terms before using a MLP encoder. The expansion terms can be selected from an external source/model (e.g., DocT5Query\cite{nogueira2019doc2query-t5}).
    \item \textbf{MLM}: The MLM encoder aggregates term weights over the logits produced by BERT's \textbf{M}asked \textbf{L}anguage \textbf{M}odel head. The weight for each term in the vocabulary is generated as follows:
    \begin{equation} \label{eq: mlm} w_i(t) = q(t)log\bigg(1 + \max_{j=1...L} ReLU\Big(h_j^\intercal e_i + b_i\Big)g(t_j)\bigg). 
    \end{equation} 
    The ReLU function ensures non-negative weights and can be replaced with e.g. a Softplus, which has similar properties but is differentiable everywhere. The $log$ normalization prevents some weights from getting too large. Term importance and passage quality scores are captured by $g(t_j)$ and $q(t)$, respectively.
    When present, the $g(t_j)$ and $q(t)$ functions can be modeled by an linear layer on top of contextualized embeddings of input tokens and the [CLS] token. Out of the three approaches using a MLM encoder, only one includes these functions.
    The choice of $\max$ aggregation and ReLU activation makes sparser representations and, at the same time, reduces training time as they disconnect the output from many paths in the computational graph.   
    \item \textbf{clsMLM}: This is a simplified version of the MLM encoder that only takes the logits of the [CLS] token, which is at the position $0$ of the sequence, as the output vector. Intuitively, this encoder squeezes the information of the whole sequence into a small [CLS] vector, which is then projected into an over-complete set of vocabulary bases:
    \begin{equation}
        w_i(t) = ReLU(h_0^\intercal e_i + b_i)
    \end{equation}
    where $h_0$ is the contextualized embedding of the CLS token. 
    
\squishend
These encoders are defined independent of input type (i.e., query or document). We can use a single shared encoder to encode both queries and documents or employ two separate encoders mixed-and-matched from the above list.

\subsubsection{Sparse regularizers.} Sparse regularizers control the sparsity of weight vectors, which is crucial for query processing efficiency. We describe three common regularization techniques used in learned sparse retrieval methods. 
\squishlist
    \item \textbf{FLOPs}: The FLOPs regularizer \cite{paria2019minimizing}, estimates the average number of floating-point operations needed to compute the dot product between two weight vectors by a smooth function. FLOPs is defined over a batch of $N$ sparse representations as follows:
    \begin{equation}
    FLOPs = \sum_{i=1}^{|V|}\Bar{a_i}^2 = \sum_{i=1}^{|V|} \Big(\frac{1}{N}\sum_{j=1}^N w_j^i\Big)^2 
    \end{equation}
    where $\Bar{a_i}$ is the estimated activation probability of the $i^{th}$ dimension. Intuitively, the FLOPS regularizer might lead to two side-effects: (1) it forces the weights to be small and (2) it encourages uniform activation probability across all dimension when the square sum is minimized.  
    \item \textbf{$L_p$ Norm}:  The family of $L_p$ norms has been commonly applied in machine learning to mitigate over-fitting. With LSR, $L_p$ is applied to the output vector rather than to model weights. $L_1$ and $L_2$ are two widely used norms.  
    \item \textbf{Top-K}: This is a simple pruning technique which only keeps the top-k highest weights and zeroes out the rest. This pruning can be applied at inference time as a post-processing step or at training time with the value of $k$ decreasing over time \cite{macavaney2020expansion}.
    \vspace{-5mm}
\squishend
\subsubsection{Supervision.} 
\begin{figure}[th]
    \centering
    \includegraphics[width=0.7\linewidth]{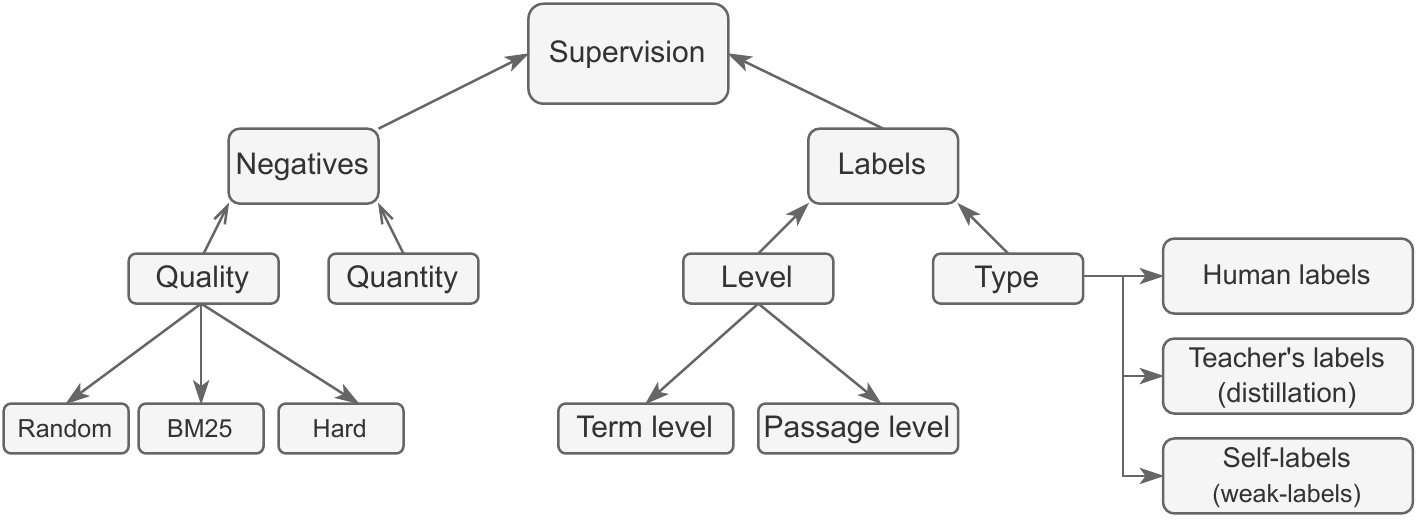}
    \caption{Aspects of supervision commonly used for learned sparse retrieval.}
    \label{fig:supervision}
    \vspace{-6mm}
\end{figure}
As some published LSR methods have identical sparse encoder(s) and sparse regularizer(s), we consider the supervision component to differentiate them and to consider its effect. 
As illustrated in Figure \ref{fig:supervision}, this supervision component is composed of two factors: negative examples and labels. 
\squishlist
    \item \textbf{Negatives:}  
    For contrastive learning, the quality and number of negatives used for training have a significant impact on performance \cite{ash2021investigating}. The more and harder the negatives, the better the result. A naive way of selecting negatives is randomly sampling non-positive passages/documents from the corpus\cite{nguyen2016ms, macavaney2020expansion, zhuang2021tilde, izacard2021unsupervised}, but this tends to create easy, less informative examples. Harder negatives can be selected from the top non-positive documents returned by BM25 or by neural retrieval models \cite{xiong2020approximate}, which can also be used to filter out false negatives \cite{reimers-2019-sentence-bert}.
    \item \textbf{Labels:} Labels for training LSR methods are classified by type and level. Types include human, teacher's, and self-labels. Human labels have good quality but are scarce and costly to collect in large quantities. Teacher's labels are generated by a previously trained model and are referred to as distillation. Self-labels or proxy-labels are generated by the model itself. Label level refers to term-level or passage/document-level labels. Term-level labels provide one score per term, while passage-level labels indicate relevance for query-passage pairs. Most methods use passage-level labels.
\squishend
\vspace{-4mm}
\subsection{Surveyed learned sparse retrieval methods}
In Table \ref{tab:lsr_methods}, we present a summary of LSR methods fit into our conceptual framework. We cover nearly all transformer-based LSR methods for text ranking in the literature\footnote{We consider the prominent doc2query document expansion methods \cite{nogueira2019document,nogueira2019doc2query-t5} in the context of pre-processing for document expansion (e.g., combined with uniCOIL), but we do not treat these as standalone \textit{retrieval} methods.}, but omit several due to time and space limitations \cite{bai2020sparterm, jang2021ultra, nair2022learning, choi2022spade}.
We group the methods into four groups by their conceptual similarity. 
We discuss how the methods fit into our framework and point out any small differences that are not described by our three components (e.g., choice of nonlinearity and including term or passage quality functions).

\paragraphHd{\textbf{A.} \label{group:A}
Methods without any expansion.} \textbf{DeepCT}\cite{dai2020context} and \textbf{uniCOIL}\cite{lin2021few} use an MLP encoder for weighting terms in queries and documents, with a slight modification to Equation \ref{eq: mlp} by removing log normalization.
Using the MLP means no expansion is applied (to query or document). DeepCT and uniCOIL only differ in supervision. DeepCT is supervised by term-recall, a term-level label defined as the ratio of relevant queries containing a term.
On the other hand, uniCOIL uses passage-level labels rather than supervising individual term scores.

\paragraphHd{\textbf{B.} \label{group:B} Methods without query expansion.} \textbf{uniCOIL$_{dT5q}$}\cite{lin2021few}, \textbf{uniCOIL$_{tilde}$}\cite{lin2021few}, and  \textbf{EPIC} \cite{macavaney2020expansion} replace the MLP document encoder in group \textbf{A} with either an $expMLP$ or $MLM$ encoder, which is capable of document expansion. As a pre-processing step, uniCOIL$_{dT5q}$ and uniCOIL$_{tilde}$ expand passages with relevant terms generated by third-party models (docT5query \cite{nogueira2019doc2query-t5}, TILDE). Instead of pre-expanding the passages, EPIC is the first to leverage the MLM architecture trained to do document expansion and term scoring end-to-end at once. On the query side, EPIC keeps the log normalization as in Equation \ref{eq: mlp}. On the document side, the ReLU in Equation \ref{eq: mlm} is replaced by a Softplus and both $q(t)$ and $g(t)$ are modeled by a linear layer with a softmax activation. 

\paragraphHd{\textbf{C.} \label{group:C} Methods without query expansion or weighting.} \textbf{DeepImpact}\cite{mallia2021learning}, \textbf{Sparta}\cite{zhao2020sparta}, \textbf{TILDE}\cite{zhuang2021tilde}, and \textbf{TILDEv2}\cite{zhuang2021fast} simplify methods in group \textbf{B} by removing the (MLP) query encoder, hence have a near-instant query encoding time but no query expansion and weighting capability. DeepImpact and TILDE$_{v2}$ can be viewed as the uniCOIL$_{dT5q}$ and uniCOIL$_{tilde}$ models without a query encoder, respectively. Sparta is simplified from EPIC by (1) removing query encoder and (2) removing $q(t)$ and $g(t_j)$ in Equation \ref{eq: mlm}. TILDE replaces the MLM head in Sparta with clsMLM.

\paragraphHd{\textbf{D.}\label{group:D} Methods with full expansion and weighting.} \textbf{Splade-max}\cite{formal2021splade} and \textbf{distilSplade-max} \cite{formal2021splade} use a shared MLM architecture on both the query and document side. The MLM enables end-to-end weighting and expansion for both query and document. Instead of selecting top-k terms as in EPIC, this Splade family uses the FLOPs regularizer during training to sparsify the representations. The difference between Splade-max and distilSplade-max is the supervision. While Splade-max is trained with multiple in-batch BM25 negatives, distilSplade-max is trained with a distillation technique using mined hard negatives. Similar to Sparta, $q(t)$ and $g(t_j)$ in Equation \ref{eq: mlm} are removed from Splade models.

\setlength{\tabcolsep}{0.4em}
\begin{table}[ht]
    \centering
    \resizebox{\textwidth}{!}{%
    \begin{tabular}{llllllll}
    \toprule \toprule
    & \multirow{2}{*}{\textbf{Method}} & \multirow{2}{*}{\textbf{Query}} & \multirow{2}{*}{\textbf{Passage}} & \multirow{2}{*}{\textbf{Reg.}} &  \multicolumn{3}{c}{\textbf{Supervision}}\\
    \cline{6-8}\vspace{-3mm}\\
    & & & & & Level & Neg. & Type \\  
    \midrule
      \multirow{2}{*}{A} & DeepCT \cite{dai2020context} & MLP & MLP & - & Term  & - & - \\
    & uniCOIL \cite{lin2021few} & MLP & MLP & - & Passage & BM25(s) & Human \\
    \midrule
     \multirow{3}{*}{B} & uniCOIL$_{dT5q}$ \cite{lin2021few} & MLP & expMLP & - & Passage & BM25(s) & Human \\
    & uniCOIL$_{tilde}$\cite{lin2021few} & MLP & expMLP & - & Passage & BM25(s) & Human\\
    & EPIC\cite{macavaney2020expansion} & MLP & MLM & Top-k  & Passage & BM25 & Human \\
    \midrule
     \multirow{4}{*}{C} & DeepImpact \cite{mallia2021learning} & BINARY & expMLP  & - & Passage & BM25 & Human \\ 
    & TILDE \cite{zhuang2021fast}& BINARY & clsMLM  & - &  Term & - & - \\ 
    & TILDEv2 \cite{zhuang2021fast}& BINARY & expMLP & - & Passage & BM25(s) & Human\\ 
    & Sparta \cite{zhao2020sparta} & BINARY & MLM & - & Passage & BM25 & Human\\
    \midrule
     \multirow{2}{*}{D} & SPLADE-max \cite{formal2021splade} & MLM & MLM & FLOPs & Passage & BM25(s) & Human \\
    & DistilSPLADE-max \cite{formal2021splade} & MLM & MLM & FLOPs & Passage &  Hard & Teacher \\
    \bottomrule
    \end{tabular}}
    \caption{Definition of existing LSR methods.
    An (s) indicates multiple negatives.}
    \label{tab:lsr_methods}
    \vspace{-8mm}
\end{table}
\vspace{-10mm}

\section{Experimental settings}
\vspace{-2mm}
For all experiments, we use Huggingface's BERT implementation with \textit{distilbert-base-cased} \cite{wolf2020transformers, sanh2019distilbert}. We train our models on the MSMarco\cite{nguyen2016ms} and TripClick datasets\cite{rekabsaz2021tripclick}. 
For models that need hard negative mining and distillation on MSMarco, we use the data provided by SentenceTransformers\footnote{\url{huggingface.co/datasets/sentence-transformers/msmarco-hard-negatives}} \cite{reimers-2019-sentence-bert} for training. For TripClick, we use the training triples\footnote{\url{github.com/sebastian-hofstaetter/tripclick}} created by \cite{hofstatter2022establishing}. We evaluate methods with the benchmarks' standard metrics, including MRR@10, NDCG@10, and Recall@1000.
In the following sections, we remove the cut-off @K for brevity.

We measure encoding latency on an AMD EPYC 7702 CPU and Tesla V100 GPU. We use a modified version of Anserini \cite{lin2016toward} for indexing passages and measure retrieval latency on an AMD EPYC 7702 CPU using 60 threads. 
For \textbf{RQ1}, we followed the same hyper-parameters and losses described in the original papers to reproduce LSR methods. For \textbf{RQ2} and \textbf{RQ3}, we train all methods on a single A100 GPU using the above mined hard negatives, and distillation data for MSMarco or the BM25 triplets for TripClick. Our Github repository contains the full configurations for all experiments.
\section{Results and analysis}
In this section we consider our three RQs. We first reproduce LSR methods in their original experimental settings (RQ1), before training them in a common setting (RQ2) and analyzing the impact of architectural differences (RQ3).

\label{sec:result}
\subsection{RQ1: Are the results from LSR papers reproducible?} 
\setlength{\tabcolsep}{0.4em}
\captionsetup[table]{skip=5pt}
\begin{table}[t]
    \centering
    \small
    \begin{tabular}{llccr}
    \toprule
    \toprule
        & \textbf{Method}   & \textbf{Original MRR} & \textbf{Reproduced MRR}  & \textbf{$\Delta$ \%} \\
    \midrule
     \multirow{2}{*}{A} & DeepCT &  24.3 & 24.6 & 1.234\\
    & uniCOIL  &  31.5 & 31.6 & 0.317\\ 
    \midrule
  \multirow{3}{*}{B} & uniCOIL$_{dT5q}$ & 35.2 & 34.7 & -1.420 \\ 
    & uniCOIL$_{tilde}$ & 34.9 & 34.8 & -0.286\\
    & EPIC$^*_{top1000}$ & 27.3 & 28.8 & 5.495 \\ 
    \midrule
    \multirow{3}{*}{C} & DeepImpact & 32.6 &  31.2 & -4.294 \\ 
    & TILDE$_{v2}^*$ & 33.3 & 33.7 & 1.201  \\ 
    & Sparta &  - & 31.0 & - \\ 
    \midrule
    \multirow{2}{*}{D} & Splade$_{max}$ & 34.0 & 34.0 & 0.000 \\ 
    & distilSplade$_{max}$ & 36.9 & 37.9 & 2.439\\ 
    \bottomrule
    \end{tabular}
    \caption{Reproduced MRR@10 scores on MSMarco dev. ($^*$) Indicates reranking results on BM25 top-1000 passages (following the original work).}
    \label{tab:reproduction}
    \vspace{-8mm}
\end{table}
We train the LSR methods using a similar experimental setup described in the original papers and code. 
The reproduced results are reported in Table \ref{tab:reproduction}. For most of the methods, we obtain scores that are slightly higher or comparable to the original work.
A slightly higher MRR was observed for DeepCT, uniCOIL, EPIC, TILDE$_{v2}$, and distilSplade$_{max}$, while DeepImpact and uniCOIL$_{dT5q}$ received slightly lower reproduced scores. Sparta was not evaluated on MSMarco in the original paper, so there is no comparison point for our result.

These reproduced results show that DeepCT and uniCOIL (without docT5query expansion) tend to be the least effective approaches, whereas distilSplade$_{max}$ achieves the highest MRR.  
Interestingly, we observe pairs of methods that have identical architectures, but different training recipes lead to a significant discrepancy in scores. uniCOIL changes the supervision signal of DeepCT from token-level weights to passage-level relevance, making a 28\% jump in MRR from $24.6$ to $31.6$. Apparently, the supervision matters a lot here; using the passage-level labels allows the model to learn the term weights more optimally for passage-level relevance. Similarly, using mined hard negatives and distillation boosts MRR from 34.0 to 37.9 with the Splade model. This change of supervision makes distilSplade$_{max}$ the most effective LSR method considered. Without this advanced training, Splade$_{max}$ performs comparably to uniCOIL$_{dT5q}$ and uniCOIL$_{tilde}$.
Looking closely at the group (\textbf{B}), EPIC seems to perform under its full capacity because it achieves a MRR substantially below the two uniCOIL variants. This may be due to the fact that EPIC was originally trained on 40000 triples, whereas the other methods were trained on up to millions of samples.

\subsection{RQ2: How do LSR methods perform with recent advanced training techniques?}
\setlength{\tabcolsep}{0.2em}
\begin{table*}[t]
    \centering
    \small
    \resizebox{\textwidth}{!}{%
    \begin{tabular}{llllcccccc|cc}
    \toprule
    \toprule
     & \multirow{2}{*}{\textbf{Method}} & \multicolumn{2}{c}{\textbf{MSMarco}} & \multicolumn{2}{c}{\textbf{DL-2019}} & \multicolumn{2}{c}{\textbf{DL-2020}} & \textbf{Index} & \textbf{RL} & \multicolumn{2}{c}{\textbf{BM25 Negs}}\\
    & & \scriptsize{MRR} & \scriptsize{R} & \scriptsize{NDCG} & \scriptsize{R} & \scriptsize{NDCG} & \scriptsize{R} & \scriptsize{GB} & \scriptsize{ms} & \multicolumn{1}{c}{\scriptsize{MRR}} & \multicolumn{1}{c}{\scriptsize{R}}   \\ 
    \midrule
    A & uniCOIL &  $27.3_{***}^{\dagger\dagger\dagger}$ & $88.0_{***}^{\dagger\dagger\dagger}$ & 59.3 & 72.9 & 54.3 & 77.9 & \textbf{1.1} &  \textbf{6.1} & 32.1 & 92.6\\ 
    \midrule
    \multirow{3}{*}{B} & uniCOIL$_{dT5q}$ &  $35.0_{*}^{\dagger}$ & 95.7 & 65.9 & 81.0 & 68.4 & 84.6 & 1.8 & 12.7 & 34.7 & 96.4\\ 
    & uniCOIL$_{tilde}$ & $36.1_{***}^{\dagger\dagger\dagger}$ & $96.8_{**}^{\dagger\dagger}$ & 69.1 & 82.2 & 69.4 & 85.2 & 2.6 & \underline{7.1} & 34.8 & \underline{96.5} \\ 
    & EPIC$_{top400}$ &  $37.2_{***}^{\dagger\dagger\dagger}$ & $97.2_{***}^{\dagger\dagger\dagger}$ & 70.9 & \underline{87.7} & \underline{71.8} & 88.7 & 9.7 & 17.7 & \textbf{35.5} & 96.4 \\
    \midrule
    \multirow{4}{*}{C} & DeepImpact &  $32.2_{**}$ & $94.7^{\dagger\dagger\dagger}$ & 63.1 & 77.2 & 63.3 & 82.1 & 1.8 & 16.1 & 32.2 & 95.4\\
    & TILDE$_{top400}$ &  $29.9^{\dagger\dagger\dagger}$ & $93.9^{\dagger\dagger\dagger}$ & 65.1 & 68.5 & 63.0 & 69.9 & 6.4  & 29.0 & 21.6 & 74.5\\ 
    & TILDE$_{v2}$ & $32.9_{**}^{\dagger\dagger}$ & 96.0 & 66.3 & 79.7 & 65.9 & 83.5 & 2.6 & 9.5 & 33.7 & 96.1\\ 
    & Sparta$_{top400}$ &  $35.3^{\dagger\dagger\dagger}$ & $96.8^{\dagger\dagger\dagger}$ & 69.1 & 81.9 & 68.1 & 85.8 & 6.1 & 26.7 & 28.3 & 88.7\\ 
    \midrule 
    \multirow{2}{*}{D} & distilSplade$_{max}$ &  \underline{$37.9^{\dagger\dagger\dagger}$} & \textbf{$98.1^{\dagger\dagger\dagger}$} & \textbf{74.8} & \textbf{87.9} & \textbf{72.5} & \textbf{89.5} & 6.3 & 122.5 & \underline{35.3} & \textbf{97.0}\\ 
    & distilSplade$_{sep}$ & \textbf{38.0} & \underline{98.0} & \underline{74.1} & \underline{87.7} & 70.6 & \underline{89.0} & 8.0  &  50.2 & - & - \\ 
    \midrule
    \multicolumn{12}{c}{\textit{***/$\dagger\dagger\dagger$ $p < 0.01$, **/$\dagger\dagger$ $p < 0.05$, */$\dagger$ $p < 0.1$ with paired two-tailed t-test}} \\ 
\multicolumn{12}{c}{\textit{Comparing with results in Table \ref{tab:reproduction} (*) and BM25 negatives results ($\dagger$)}} \\ 
    \bottomrule
    \end{tabular}}
    \caption{Results with cross-encoder distillation on hard negatives (left) and BM25 negatives on MS MARCO (two rightmost columns). \textbf{RL} indicates the latency (ms/q) for query encoding and retrieval.}
    \label{tab:lsr_distil}
    \vspace{-8mm}
\end{table*}
Variations in environments, as shown in RQ1, make it difficult to fairly compare LSR methods and can lead to inaccurate conclusions. To eliminate these discrepancies, we train all methods in a consistent environment, which we show to be effective in this section. We focus on the most effective supervision setup, which is distilSplade$_{max}$ trained using distillation and hard negatives. Table \ref{tab:lsr_distil} shows the results of the LSR methods under this setting. Note that several methods (DeepCT and uniCOIL; Splade variants) will have identical scores in this experiment as they collapse into the same model. We only report a representative method in these cases. 

Comparing to the results of \textbf{RQ1} (Table \ref{tab:reproduction}), we find that the least effective methods (DeepCT, now equivalent to uniCOIL) and the most effective method (distilSplade$_{max}$) remain in the same positions. Methods between these two endpoints move around with substantial changes in their effectiveness. Out of 10 methods we reproduced in Table \ref{tab:reproduction}, we observe an upward trend on seven methods, while the remaining three methods stay the same or perform worse. The biggest jumps are seen using EPIC and Sparta, with a relative improvement of 8.0 and 4.2 MRR points on MSMarco, respectively.
The increase in EPIC's effectiveness, which is due to the combination of longer training time and improved supervision, moves the approach's relative ranking from the second worst to the second best, with metrics competitive with distilSplade$_{max}$ on MSMarco. On TREC DL 2019 and TREC DL 2020, the gap in NDCG@10 between EPIC and distilSplade$_{max}$ is higher. The increased MRR@10 on MSMarco also brings Sparta a nice efficiency-effectiveness trade-off: since there is no query encoder with Sparta, there is no need for a GPU at retrieval time.

In addition to EPIC and Sparta, we also observe positive trends with DeepCT, DeepImpact, uniCOIL$_{dT5q}$ and uniCOIL$_{tilde}$; however, the change is relatively marginal. We observe decreased effectiveness on uniCOIL and TILDE$_{v2}$. While the decline with TILDE$_{v2}$ is small, the drop with uniCOIL (32.1$\rightarrow$27.3) is quite large. Indeed, without expansion capability, no soft-matching could be possible, which renders a challenge for uniCOIL to reconstruct the MarginMSE's loss margin produced by a cross-encoder teacher, which is capable of soft-matching.

Regarding architecture types, methods using the MLM architecture, either on the document or query side (EPIC, Sparta, Splade), generally perform better than those using other architectures (clsMLM, MLP, expMLP, BINARY) on all three datasets. However, MLM also increases index size and latency significantly. For instance, EPIC's index is at least 6 GB larger than other methods in the group. Notably, distilSplade$_{max}$ not only creates a large index but also has a notably high retrieval latency, almost 20 times slower than the fastest method.

The latency issue in Splade is related to using the same shared MLM encoder for query and documents, resulting in similar term activation probability between queries and documents. We confirmed this by replacing the shared encoder with two separate ones (distilSplade$_{sep}$), which reduced latency from 122.5 ms to 50.2 ms, a 59\% decrease. This benefit of separate encoders was also reported in \cite{lassance2022efficiency}, and our results further support its substantial impact.


In the last two columns of Table \ref{tab:lsr_distil}, we provide additional MSMarco results with training using BM25 in-batch negatives (the same as uniCOIL's original setup). We find that using hard negatives with distillation is generally more effective than using BM25 negatives, though not with uniCOIL or TILDE$_{v2}$.

\vspace{-3mm}
\subsection{RQ3: How does the choice of encoder architecture and regularization affect results?}
\begin{table}[h]
 \centering
    \resizebox{\textwidth}{!}{%
    \begin{tabular}{llllllllll}
    \toprule
    \toprule
    \multirow{2}{*}{\textbf{Effect}} & & \multirow{2}{*}{\textbf{Control}} & \multirow{2}{*}{\textbf{Change}} & \multicolumn{2}{c}{\textbf{MSMarco}} & \multicolumn{1}{c}{\textbf{DL 2019}} & \multicolumn{1}{c}{\textbf{DL 2020}} & \multicolumn{1}{c}{\textbf{Latency}} & \multicolumn{1}{c}{\textbf{Index}} \\
    & & & & \multicolumn{1}{c}{\scriptsize{MRR}} & \multicolumn{1}{c}{\scriptsize{R}} & \multicolumn{1}{c}{\scriptsize{NDCG}} & \multicolumn{1}{c}{\scriptsize{NDCG}} & \multicolumn{1}{c}{\scriptsize{ms}} & \multicolumn{1}{c}{\scriptsize{GB}} \\ 
    \midrule
     \multirow{2}{*}{\textbf{Doc weighting}} & $1_a$ & $Q_{MLM}$ & $D_{BIN} \rightarrow D_{MLP}$ & 16.7\Plus{18.3} & 86.0\Plus{11.0} & 44.1\Plus{26.8} & 42.9\Plus{24.5} & 11.4\Plus{04.2} & 0.6\Plus{0.7}\\
      & $1_b$ & $Q_{MLP}$ & $D_{eBIN} \rightarrow D_{eMLP}$ & 08.2\Plus{27.9} & 76.2\Plus{20.6} & 30.4\Plus{38.7} & 27.5\Plus{41.9} & 10.8\Minus{03.7} & 1.2\Plus{1.4} \\
      \midrule
     \multirow{2}{*}{\textbf{Query weighting}} &$2_a$& $D_{eMLP}$ & $Q_{BIN}  \rightarrow Q_{MLP}$ & 32.9\Plus{3.2} & 96.0\Plus{0.8} & 66.3\Plus{2.8} & 65.9\Plus{3.5} & 09.5\Minus{0.9} &  2.6\Plus{0.0}\\ 
      & $2_b$ & $D_{MLM}$ & $Q_{BIN}  \rightarrow Q_{MLP}$ & 35.2\Plus{1.9} & 96.5\Plus{0.7} & 69.4\Plus{1.5} & 69.7\Plus{2.1} & 28.9\Minus{7.9} & 8.6\Plus{1.1} \\ 
     \midrule
     \multirow{3}{*}{\textbf{Doc expansion}}  & $3_a$ & $Q_{MLM}$ & $D_{MLP} \rightarrow D_{MLM}$ & 34.9\Plus{3.1} & 97.0\Plus{0.9} & 70.9\Plus{3.3} & 67.4\Plus{3.2} & 15.6\Plus{34.6} & 1.3\Plus{6.7} \\
      & $3_b$ & $Q_{MLP}$ & $D_{MLP}  \rightarrow D_{MLM}$ & 27.5\Plus{10.0} & 89.7\Plus{8.2} & 59.3\Plus{12.0} &  54.3\Plus{17.9} & 27.5\Plus{10.5} & 1.2\Plus{6.9} \\ 
     \midrule 
     \multirow{2}{*}{\textbf{Query expansion}} & $4_a$ & $D_{MLM}$ & $Q_{MLP}  \rightarrow Q_{MLM}$ & 38.0\Plus{0.0} & 97.0\Plus{0.1} & 71.3\Plus{2.8} & 72.1\Minus{1.3} & 12.9\Plus{37.3}& 8.0\Minus{0.1}\\  
      &$4_b$ & $D_{MLP}$ & $Q_{MLP}  \rightarrow Q_{MLM}$ & 27.5\Plus{7.5} & 89.7\Plus{7.4} & 59.3\Plus{11.6} & 54.3\Plus{13.1} & 06.1\Plus{9.5} & 1.2\Plus{0.1} \\  
     \midrule
    \textbf{Regularization} & $5_a$ & \makecell[l]{$Q_{MLP}$ \\ $D_{MLM}$} & $FLOPs \rightarrow Topk$ & 38.0\Plus{0.0} & 97.9\Minus{0.3} & 71.3\Plus{0.8} & 72.1\Plus{0.1} & 12.8\Plus{4.3} & 8.1\Minus{0.7}   \\
    \midrule 
    & & & & \multicolumn{6}{c}{\textbf{TripClick}}\\
    \cline{5 -10}\vspace{-3mm}\\
    \multirow{2}{*}{\textbf{}} & & \multirow{2}{*}{\textbf{}} & \multirow{2}{*}{\textbf{}} & \multicolumn{2}{c}{\textbf{HEAD(dctr)}} & \multicolumn{1}{c}{\textbf{TORSO(raw)}} & \multicolumn{1}{c}{\textbf{TAIL(raw)}} & \multicolumn{1}{c}{\textbf{Latency}} & \multicolumn{1}{c}{\textbf{Index}} \\
    & & & & \multicolumn{1}{c}{\scriptsize{NDCG}} & \multicolumn{1}{c}{\scriptsize{R}} & \multicolumn{1}{c}{\scriptsize{NDCG}} & \multicolumn{1}{c}{\scriptsize{NDCG}} & \multicolumn{1}{c}{\scriptsize{ms}} & \multicolumn{1}{c}{\scriptsize{GB}} \\ 
    \midrule
     \multirow{2}{*}{\textbf{Doc weighting}} & $1_a$ & $Q_{MLP}$ & $D_{BIN} \rightarrow D_{MLP}$ & 6.5\Plus{18.9} & 69.7\Plus{18.4} & 10.7\Plus{17.5} & 16.2\Plus{13.2} & 2.0\Minus{0.1} & 0.3\Plus{0.3}\\
      & $1_b$ & $Q_{MLP}$ & $D_{eBIN} \rightarrow D_{eMLP}$ & 5.7\Plus{21.0} & 67.2\Plus{21.1} & 9.1\Plus{20.4} & 13.9\Plus{16.5} & 2.5\Minus{0.3} & 0.4\Plus{0.5} \\
      \midrule
     \multirow{2}{*}{\textbf{Query weighting}} &$2_a$& $D_{MLM}$ & $Q_{BIN}  \rightarrow Q_{MLP}$ & 26.3\Plus{3.9} & 90.0\Plus{1.9} & 31.3\Plus{3.3} & 34.2\Plus{3.8} & 3.2\Minus{0.0} &  1.8\Minus{0.1}\\ 
      & $2_b$ & $D_{MLP}$ & $Q_{BIN}  \rightarrow Q_{MLP}$ & 24.2\Plus{1.1} & 87.3\Plus{0.8} & 27.7\Plus{0.4} & 29.4\Plus{0.0} & 2.1\Minus{0.2} & 0.5\Plus{0.1} \\ 
     \midrule
     \multirow{3}{*}{\textbf{Doc expansion}}  & $3_a$ & $Q_{MLM}$ & $D_{MLP} \rightarrow D_{MLM}$ & 27.9\Plus{2.2} & 90.9\Plus{1.0} & 32.7\Plus{1.5} & 34.1\Plus{3.9} & 4.6\Plus{1.6} & 0.7\Plus{0.7} \\
      & $3_b$ & $Q_{MLP}$ & $D_{MLP} \rightarrow D_{MLM}$ & 25.3\Plus{4.7} & 88.1\Plus{3.7} & 28.2\Plus{6.1} & 29.4\Plus{7.9} & 1.9\Plus{1.6} & 0.6\Plus{0.8} \\ 
     \midrule 
     \multirow{2}{*}{\textbf{Query expansion}} & $4_a$ & $D_{MLM}$ & $Q_{MLP}  \rightarrow Q_{MLM}$ & 30.0\Plus{0.1} & 91.8\Plus{0.1} & 34.2\Minus{0.1} & 37.4\Plus{0.6} & 3.4\Plus{2.8}& 1.4\Minus{0.0}\\  
      &$4_b$ & $D_{MLP}$ & $Q_{MLP}  \rightarrow Q_{MLM}$ & 25.3\Plus{2.6} & 88.1\Plus{2.8} & 28.2\Plus{4.5} & 29.4\Plus{4.6} & 1.9\Plus{2.7} & 0.6\Plus{0.0} \\  
     \midrule
    \textbf{Regularization} & $5_a$ & \makecell[l]{$Q_{MLP}$ \\ $D_{MLM}$} & $L1 \rightarrow Topk$ & 30.0\Plus{0.1} & 91.8\Plus{0.1} & 34.2\Plus{0.3} & 37.4\Plus{0.7} & 3.4\Minus{0.2} & 1.4\Plus{0.3}   \\
    \bottomrule
    \end{tabular}}
     \caption{The effects of architecture and regularizer on MSMarco and TripClick.\\ \textit{We use names that better reflect the architectural differences between methods.\\ Visit our \href{https://github.com/thongnt99/learned-sparse-retrieval}{Github repository} to see the full configurations and original names.}}
     \label{tab:ablation}
    \vspace{-5mm}
\end{table}

In this RQ, we aim to quantify how different factors (\textit{query expansion}, \textit{document expansion}, \textit{query weighting}, \textit{document weighting}, \textit{regularization}) affect the effectiveness and efficiency of LSR systems. To eliminate potential confounding factors due to minor differences between groups (e.g., choice of nonlinearity), we perform a series of controlled experiments in which we make single architectural changes while holding the rest of the architecture constant.

In  Table \ref{tab:ablation}, numbers before \Plus{} or \Minus{} are the metrics before a change (left side of arrow),  while numbers after these symbols show the effect of a change (right). We see that document weighting seems to be the most crucial component since the systems without this component fail on all three datasets. In row $1_{(a,b)}$, the system with a binary document encoder shows very low MRR and NDCG scores regardless of MLM or MLP on the query side. On both MSMarco and TripClick,  enabling document weighting (by replacing the binary document encoder with an MLP) improves the effectiveness by a large margin (at least 11 points) with reasonable latency and index size increases. 
Without document weighting, the models are not able to identify important terms in documents.

Similarly, as shown in rows $2_{(a,b)}$, we control the document side and change the binary query encoder to an MLP query encoder to observe the effect of query weighting. The result suggests that query weighting has a moderate contribution to the ranking metrics overall. Still, interestingly, it causes almost no harm to the index size or even reduces the latency. Note that the latency of the MLP query encoder here is measured on GPU; therefore, the encoding overhead is tiny. The improved overall latency is mostly due to the MLP reducing the weights of some non-useful query terms to zero, making queries shorter. The effect is quite consistent between MSMarco and TripClick collections.

Regarding the expansion factors, we observe the cancellation effect between query expansion and document expansion. Indeed, with the absence of expansion on one side ($3_{b}$: $Q_{MLP}$ has no query expansion, $4_b$: $D_{MLP}$ has no document expansion), the expansion on the other side largely improves the ranking metrics with at least $7.4 (2.6)$ points and at most $17.9 (7.9)$ points overall on MSMarco (TripClick). The cost of latency, in this case, is rather low. 
The numbers in rows $3_a$ and $4_a$ indicate that query and document expansion have a cancellation effect. That is, query expansion reduces the benefit of performing document expansion and vice versa. Row $4_a$ shows that when document expansion is in place, query expansion has minimal impact on ranking effectiveness and incurs a relatively high latency overhead (increases of 289\% and 82\% on MSMarco and TripClick). Row $3_a$ shows a similar trend, with document expansion making moderate contributions to system effectiveness. 
On TripClick, the cancellation interaction between the two factors is less strong. Overall, this cancellation effect suggests that including both expansion components may not be necessary.

Lastly, to examine the effect of regularization, we keep the model's architecture constant and change the FLOPs/L$_1$ regularizer during training to Topk pruning during inference. As shown in rows $5_a$, changing the regularization approach does not significantly affect effectiveness or efficiency.

\setlength{\tabcolsep}{0.2em}
\begin{table}[h]
    \centering
    \small
    \begin{tabular}{lcccc|cccc}
    \toprule
    \toprule
     \multirow{2}{*}{\textbf{Method}} & \multicolumn{4}{c}{\textbf{MSMarco-dev}} &  \multicolumn{4}{c}{\textbf{TripClick-HEAD(dctr)}}\\
     \cline{2-5} \cline{6-9} 
    & \scriptsize{MRR} & \scriptsize{R} & \scriptsize{Index(GB)} & \scriptsize{RL(ms)}     & \scriptsize{NDCG} & \scriptsize{R} & \scriptsize{Index(GB)} & \scriptsize{RL(ms)} \\ 
    \midrule
    distilSplade$_{sep}$ & 38.0 & 98.0  & 8.0 & 50.2 & 30.1 & 91.9 & 1.4 & 6.3 \\ 
    distilSplade$_{qMLP}$ & 38.0 & 97.9 & 8.1 & 12.9 & 30.0 & 91.8 & 1.4 & 3.4 \\
    distilSplade$_{dMLP}$ & 34.9$^{*}$ & 97.0$^{*}$ & 1.3 & 15.6 & 27.9$^{*}$ & 90.9$^{*}$ & 0.7 & 4.6\\
    \bottomrule
    \end{tabular}
    \caption{Results with only query expansion or only document expansion.\\\textit{* $p < 0.01$ with paired two-tailed t-test}}
    \label{tab:lsr_no_expansion}
    \vspace{-8mm}
\end{table}


In Table \ref{tab:lsr_no_expansion}, we show the results of systems with expansion only on either the query or the document side. In the table, distilSplade$_{qMLP}$ denotes the distilSplade$_{sep}$ with the MLM query encoder replaced by an MLP query encoder; hence no query expansion is involved. Similar interpretation applies for distilSplade$_{dMLP}$. As can be seen, distilSplade$_{qMLP}$ makes no significant changes on ranking metrics, while reducing the retrieval latency by more than 74\% and 46\% on MSMarco and TripClick, respectively. distilSplade$_{dMLP}$ exhibits a similar latency improvement, but suffers from a significant drop in effectiveness. 
In practice, distilSplade$_{qMLP}$ could be viewed 
as a more efficient drop-in replacement for the full model. This use of $qMLP$ is complementary to other changes (e.g., using a smaller encoder as in \cite{lassance2022efficiency}) to improve the efficiency of LSR.

\setlength{\tabcolsep}{0.3em}

\section{Conclusion}
In this work, we introduced a conceptual framework for learned sparse retrieval that unifies existing LSR methods under the perspective of three components.
After reproducing these methods, we carried out a series of experiments to isolate the effect of single changes on a model's performance.
This analysis led to several findings about the components, including that we can remove the query expansion from a SOTA system, leading to a significant latency 
improvement without compromising the system's effectiveness. 
While this study covered the most prominent transformer-based LSR methods, several others could not be considered due to time and computing constraints (e.g., \cite{jang2021ultra, bai2020sparterm, nair2022learning, choi2022spade}).
We plan to incorporate them into our implementation as future work.

\section*{Acknowledgement}
We thank Maurits Bleeker from the UvA IRLab for his feedback on the paper. 
\bibliographystyle{splncs04}
\bibliography{main}
\end{document}